\documentclass[lettersize,journal]{IEEEtran} 

\usepackage{color,soul}
\usepackage{subcaption}
\usepackage{graphicx}      
\usepackage{comment}
\usepackage{amsmath,amssymb,amsfonts}
\usepackage{tikz}

\title{\LARGE \bf
	Resource-Aware Stealthy Attacks in Vehicle Platoons
}
\author{Ali Eslami$^{1}$ and Mohammad Pirani$^{2}$
	\thanks{$^{1}$Ali Eslami is with the Department of Electrical and Computer Engineering, Concordia University, Montreal, Quebec H3G 1M8, Canada.
		{\tt\small a\_esla@live.concordia.ca}}
	\thanks{$^{2}$Mohammad Pirani is with the Department of Mechanical Engineering, University of Ottawa, Ottawa, ON K1N 6N5, Canada. 
		{\tt\small mpirani@uottawa.ca}	
		}
	\thanks{This work has been submitted to the IEEE for possible publication. Copyright may be transferred without notice, after which this version may no longer be accessible.}
}
\begin{document}
\maketitle
\thispagestyle{plain}
\pagestyle{plain}
	

	\begin{abstract}                
		Connected and Autonomous Vehicles (CAVs) are transforming modern transportation by enabling cooperative applications such as vehicle platooning, where multiple vehicles travel in close formation to improve efficiency and safety. However, the heavy reliance on inter-vehicle communication makes platoons highly susceptible to attacks, where even subtle manipulations can escalate into severe physical consequences. While existing research has largely focused on defending against attacks, far less attention has been given to stealthy adversaries that aim to covertly manipulate platoon behavior. This paper introduces a new perspective on the attack design problem by demonstrating how attackers can guide platoons toward their own desired trajectories while remaining undetected. We outline conditions under which such attacks are feasible, analyze their dependence on communication topologies and control protocols, and investigate the resources required by the attacker. By characterizing the resources needed to launch stealthy attacks, we address system vulnerabilities and informing the design of resilient countermeasures. Our findings reveal critical weaknesses in current platoon architectures and anomaly detection mechanisms and provide methods to develop more secure and trustworthy CAV systems.
	\end{abstract}

	
\section{Introduction}

\subsection{Motivation and Context}

The advent of Connected and Autonomous Vehicles (CAVs) is transforming the landscape of modern transportation. By integrating advanced sensing, computation, and vehicle-to-everything (V2X) communication, CAVs are envisioned to enhance traffic efficiency, reduce fuel consumption, and improve road safety through cooperative and automated driving \cite{sun2021survey}. The connectivity of autonomous vehicles enables mission-critical applications such as cooperative collision avoidance, dynamic map sharing, and vehicle platooning \cite{shiwakoti2020investigating}.

Among the many CAV applications, vehicle platooning has emerged as a promising yet particularly vulnerable use case. In a platoon, multiple vehicles travel in close formation under coordinated control, exchanging real-time information to maintain small inter-vehicle distances and ensure string stability \cite{jia2015survey}. While this approach yields substantial gains in traffic throughput and fuel efficiency, similar to other applications of CAVs, it is vulnerable to different attacks \cite{sedar2023comprehensive}. Studies have documented a wide range of attacks in vehicular networks, including False Data Injection (FDI), replay, denial-of-service (DoS), Sybil attacks, GPS spoofing, and malware injection \cite{sedar2023comprehensive}. The safety–security coupling in CAVs implies that even subtle attacks can escalate into physical accidents with severe consequences \cite{ju2022survey}. That indicates that compromising a single communication channel might destabilize the entire platoon. 

Due to the importance of addressing cybersecurity concerns in vehicle platoons, our objective in this paper is to investigate the possibility and design procedure of attacks in vehicle platoons. Our aim is to perform a vulnerability assessment and investigate the required resources that are necessary for the attacker to design attack signals capable of manipulating the platoon behavior.

\subsection{Related Work}
\label{subsec:Related Work}
Vulnerabilities and resiliency of vehicle platooning have become an avenue of research in recent years \cite{ding2024distributed,han2024resilient,zhao2024safeguard}. To investigate the vulnerabilities of a single vehicle, several attack types have been investigated in the literature including Zero Dynamics Attacks (ZDA) \cite{teixeira2012attack}, Pole Dynamics Attack (PDA) \cite{du2024stealthy}, Replay Attacks \cite{ye2019stochastic}, and Covert Attacks \cite{barboni2020detection}. 

The resiliency of vehicle platoons can be categorized broadly into two domains: resiliency by design and control input modification. The first category, resiliency by design, relies primarily on the Mean-Subsequence-Reduced (MSR) approach \cite{leblanc2013resilient}, which requires the communication graph to be sufficiently \textit{r-robust} relative to the maximum number of malicious neighbor nodes. By progressively shrinking a convex set formed from neighbors' values, consensus among healthy agents is eventually achieved. This idea has been utilized in different settings, such as asynchronous Multi-Agent System (MAS) \cite{dibaji2017resilient}, MAS with delayed network \cite{wu2017secure}, and attitude consensus \cite{rezaee2019resilient}. One can refer to \cite{pirani2022impact} for a thorough analysis of the impact of different network topologies on platoon resiliency. However, this approach is inherently conservative, often demanding graph connectivity conditions that may not exist in practical platoon topologies, prompting recent works to relax these conditions for leader-follower scenarios from \textit{$2f+1$-robust} in the whole network to \textit{$f+1$-robust} in the healthy agents \cite{liao2025leader}.
In the control input modification approach (which has been investigated mainly against False Data Injection (FDI) Attacks), the attack signals are estimated and utilized in the controller of the agents. This idea has been investigated in different settings \cite{meng2020adaptive,huo2022secure,sun2024adaptive}.

In addition to these approaches, attack recovery methods have been developed to restore system performance once an attack is detected. For example, \cite{130,131} employ a simplified platoon model with acceleration inputs to address actuator attacks on the leader or last vehicle. Safety is evaluated via reachable set analysis, and when necessary, additional input constraints are applied to avoid collisions; these constraints are removed during normal operation. While effective in specific scenarios, these methods are limited by the assumption that attacks occur only on the leader or the last vehicle, reducing their applicability to broader threat models. For a comprehensive overview of CAV resiliency research, the interested reader is also referred to \cite{ju2022survey} and the references therein.

While resilient control of vehicle platoons has been investigated in the literature, significantly less attention has been paid to the attack design problem from the attacker's perspective. The importance of this analysis lies in the fact that it can act as a vulnerability assessment of the system, providing us with the knowledge of the resources required by the attackers. The authors in \cite{taheri2020undetectable} investigated undetectable attacks in MAS by analyzing how an attacker can prevent consensus while remaining stealthy from a distributed residual generation mechanism. In \cite{sun2024adaptive}, the authors considered FDI attacks on communication between agents where stealthiness is defined based on the probability distribution of the distributed residual. Moreover, \cite{zhang2024ripple} investigated the required graph and system conditions for stealthy FDI attacks that lead to unbounded consensus error, while \cite{luo2024submodularity} proposed an optimization problem to maximize the divergence of under-attacked agents with bounded actuator attacks. Finally, \cite{tsang2020sparse} investigated scenarios where, given a certain number of under-attacked links, the goal is to minimize the attack signal value to make the consensus error dynamics unstable.
\textit{Critical Research Gap}: Existing attack research mainly aims to destabilize MAS or induce unbounded tracking errors. This focus overlooks a more subtle and potentially dangerous class of attacks, where adversaries manipulate the platoon into following a desired trajectory of their choosing. Such attacks represent a shift from sabotage to sophisticated manipulation, enabling covert control of platoon and vehicle behavior.

\subsection{Research Objectives and Approach}

This paper addresses the identified gap by developing a comprehensive framework for stealthy attacks against leader-follower vehicle platoons, where the attacker's objective is to steer either the entire platoon or a subset of it toward tracking the attacker's own state signal rather than the legitimate leader of the platoon. This perspective shifts the goal of the attacker from sabotage to subtle manipulation, representing a more covert and potentially damaging class of attacks (e.g., undetected trajectory shifts could mislead entire fleets into wrong highways, cause coordinated traffic jams, or unsafe but seemingly valid routes.).

To study the feasibility and impact of such attacks, our approach addresses two core challenges: (i) identifying which communication links or vehicles must be compromised to achieve the attacker's goal, and (ii) designing attack signals that steer the system toward the attacker's state while remaining stealthy. Our approach investigates this threat across multiple scenarios, including when the communication channels of the leader are vulnerable to attacks and when they are secure. Furthermore, we also investigate the effects of the control input protocol on the possibility of attacks. Finally, we also discuss the required resources for the attacker in each scenario and provide ideas on designing secure and resilient vehicle platoons against attacks.

Therefore, the contributions of this paper can be summarized as follows:

\begin{enumerate}
	\item We propose a new stealthy attack strategy for leader–follower vehicle platoons, in which the attacker seeks to manipulate the platoon to track its own state rather than merely destabilizing the formation.
	\item Attack signals are designed to achieve the attacker's objectives under various scenarios—both when the outgoing communication channels of the leader are vulnerable to attacks and when they are secure. The conditions for maintaining stealthiness are rigorously analyzed in each scenario.
	\item The influence of the agents’ control protocol on the attacker's degradation capabilities is examined, demonstrating how the control protocol can inherently restrict the feasible set of stealthy attacks.
	\item The required resources for the attacker are investigated and potential ideas and concepts on designing resilient vehicle platoons are provided.
\end{enumerate}
The remainder of this paper is structured as follows: Section \ref{sec:System Model} presents the system model, platoon topologies, and attack model foundations. Section \ref{sec:DynamicController} develops attack strategies for distributed dynamic controllers, while Section \ref{sec:StaticController} addresses static controller scenarios. Section \ref{sec:Discussions} discuss the implications for designing resilient vehicle platoons and Section \ref{sec:Numerical Example} provides comprehensive numerical case studies demonstrating attack effectiveness and limitations. Finally, we will conclude this paper in Section \ref{sec:Conclusion}.
	\section{System Model \& Preliminaries}
	\label{sec:System Model}
	\subsection{Platoon Topologies \& Graph Preliminaries}
	Each vehicle in the platoon can be modeled as a node in a directed graph. Let \(\mathcal{G}_s = (\mathcal{V}_s, \mathcal{E}_s, \mathcal{A}_s)\) represent the communication graph capturing information flow among the \(N\) follower vehicles. The set of follower vehicles is denoted by \(\mathcal{V}_s = \{1, \ldots, N\}\). The edge set \(\mathcal{E}_s \subseteq \mathcal{V}_s \times \mathcal{V}_s\) contains ordered pairs \((j, i)\), indicating a directed link from vehicle \(j\) to vehicle \(i\). The adjacency matrix is given by \(\mathcal{A}_s = [a_{ij}^{s}] \in \mathbb{R}^{N \times N}\), where \(a_{ij}^{s} = 1\) if vehicle \(j\) can transmit information to vehicle \(i\), and \(a_{ij}^{s} = 0\) otherwise. Similarly, we denote $\mathcal{N}_i$ as the neighboring set of node $i$, i.e., the set where there exist an edge between agent $i$ and the neighboring set. We assume that no self-loops exist in the graph which indicates that \(a_{ii}^{s} = 0\) and that the graph $\mathcal{G}_s$ is bidirectional, i.e., \(a_{ij}^{s} = 1\) indicates \(a_{ji}^{s} = 1\). 
	
	We now consider the augmented graph \(\mathcal{G} = (\mathcal{V}, \mathcal{E}, \mathcal{A})\), which extends \(\mathcal{G}_s\) by including the leader vehicle (node 0). Here, \(\mathcal{V} = \{0, 1, \ldots, N\}\), \(\mathcal{E} \subseteq \mathcal{V} \times \mathcal{V}\), and \(\mathcal{A} = [a_{ij}^{s}] \in \mathbb{R}^{(N+1) \times (N+1)}\).
	
	Let \(A_0 = \text{diag}\{a_{i0}^{s}\}\) be the pinning matrix, where \(a_{i0} = 1\) indicates that the leader (vehicle 0) sends information to vehicle \(i\), and \(a_{i0}^{s} = 0\) otherwise.
	The Laplacian matrix \(\mathcal{L} = [l_{ij}] \in \mathbb{R}^{N \times N}\) is then defined by:
	\[
	l_{ij} \triangleq 
	\begin{cases}
		-a_{ij}^{s}, & \text{if } i \neq j, \\
		\sum_{j \neq i} a_{ij}^{s}, & \text{if } i = j.
	\end{cases}
	\]
	
	Since the leader has no incoming edges, the matrix \( \mathcal{L} \) is structured as:
	\begin{equation}
		\mathcal{L} =
		\begin{bmatrix}
			0 & \mathbf{0}_{1 \times N} \\
			\mathcal{L}_2 & \mathcal{L}_1
		\end{bmatrix}
	\end{equation}
	where \( \mathcal{L}_2 \in \mathbb{R}^{N \times 1} \), and \( \mathcal{L}_1 \in \mathbb{R}^{N \times N} \). Since \( \mathcal{G}_s \) is bidirectional, \( \mathcal{L}_1 \) is symmetric.
	
	There exist different classes of topologies that are popular in vehicle platoons and we have considered \textit{k-Nearest Neighbor Leader Tracking} Topology, which is a type of nearest neighbor topology where each agent is communicating bidirectionally with its k nearest neighbor and the communication channels of the leader to its k nearest neighbors are unidirectional.
	Note that the discussions of this paper can be easily extended to other platoon topologies as well.
	\subsection{System Dynamics}

Consider a platoon of vehicles traveling on a road, consisting of a leader (denoted by 0) and \(N\) following vehicles labeled \(i = 1, \ldots, N\). Let \(p_i(t)\) and \(v_i(t)\) represent the position and velocity of vehicle \(i\) at time \(t\). The nonlinear longitudinal dynamics for each vehicle are described as follows, based on the models in \cite{zheng2015stability, wen2019sampled}:

\begin{equation}
	\left\{
	\begin{aligned}
		&\frac{\eta_i}{\theta_r} T_i(t) = m_i \dot{v}_i(t) + C_i^A v_i^2(t) + m_i g f_i \\
		&\tau \dot{T}_i(t) + T_i(t) = T_i^e(t)
	\end{aligned}
	\right.
	\label{eq:vehicle_dynamics}
\end{equation}
Here, \(m_i\) is the mass of vehicle \(i\), \(C_i^A\) is the aerodynamic drag coefficient, \(g\) denotes gravitational acceleration, \(f_i\) is the rolling resistance coefficient, and \(T_i(t)\) is the actual engine torque. The desired engine torque is denoted by \(T_i^e(t)\), while \(\tau\) represents a time constant for powertrain lag. Parameters \(\theta_r\) and \(\eta_i\) are the wheel radius and mechanical efficiency, respectively.

Model \eqref{eq:vehicle_dynamics} approximates the powertrain dynamics using a first-order inertial response. To simplify analysis, many studies including \cite{zheng2015stability} apply feedback linearization techniques. This results in the following relationship for the desired torque:
\begin{align}
	\label{eq:Te}
	\begin{split}
	T_i^e(t) &= \frac{1}{\eta_i} [ C_i^A v_i(t)(2\tau \dot{v}_i(t) + v_i(t)) \\&+ m_i g f_i + m_i u_i(t) ] \theta_r
	\end{split}
\end{align}
where \(u_i(t)\) represents the control input. Substituting equation \eqref{eq:Te} into \eqref{eq:vehicle_dynamics}, we obtain the linearized model:
\begin{equation}
	\tau \dot{a}_i(t) + a_i(t) = u_i(t)
	\label{eq:linear}
\end{equation}
where \(a_i(t)\) denotes the acceleration of vehicle \(i\). For simplicity, we define the state vector of vehicle \(i\) as \(x_i(t) = [p_i(t), v_i(t), a_i(t)]^T\). Using this representation and dropping the time notation $(t)$, the vehicle dynamics in \eqref{eq:linear} can be written in the following compact state-space form:
\begin{equation}
	\dot{x}_i = A x_i + B u_i, \quad i \in \{0,...,N\}
	\label{eq:MASdynamics}
\end{equation}
where the matrices \(A\) and \(B\) are given by:
\[
A = \begin{bmatrix}
	0 & 1 & 0 \\
	0 & 0 & 1 \\
	0 & 0 & -\frac{1}{\tau}
\end{bmatrix}, \quad
B = \begin{bmatrix}
	0 \\
	0 \\
	\frac{1}{\tau}
\end{bmatrix}.
\]
	 
	The communication channels among the vehicles are modeled through a directed graph \( \mathcal{G} \), which satisfies the following assumption.
	
	\textbf{Assumption 1}: The subgraph \( \mathcal{G}_s \), representing only the followers, is bidirectional. Moreover, the full graph \( \mathcal{G} \) contains directed paths from the leader to every follower (i.e., a directed spanning tree with the leader as the root exists).
	
	This paper considers the distributed tracking control problem for the vehicle platoons, defined through the following definition:
	
	\textbf{Definition 1}: For a vehicle platoon, the followers will track the leader and follow a constant spacing policy \cite{zheng2017platooning}, if each follower's state converges to the leader’s state with a specific constant spacing, i.e.,
	\begin{equation}
		\label{eq:TrackingDefinition}	
		\lim_{t \to \infty} \|x_i(t) - x_0(t)+d_i\|_2 = 0, \ for \; all \; i = 1, \ldots, N.
	\end{equation}
	where $d_i=[id^0,0,0]$ where $d^0$ is the desired spacing between the leader vehicle and the first follower.
	 
	Different from some other works such as \cite{han2024resilient} and \cite{he2025resilient}, where the leader's input is assumed to be zero or shared with all followers, we consider the more general scenario where \( u_0 \) is nonzero, time-varying, and unknown to followers. This is addressed under the following assumption.
	
	\textbf{Assumption 2}: The leader's control input $u_0$ is continuous and bounded, i.e., $||u_0||_\infty \le \gamma$, where $\gamma$ is a positive constant. Furthermore, $(A,B)$ is stabilizable.
	
	\subsection{Distributed Controllers}
	Here we consider two different distributed controllers: 
	\begin{enumerate}
		\item Dynamic Controller
		\item Static Controller.
	\end{enumerate}
	The reasoning behind choosing different controllers is to show how the choice of the control protocol can limit the attacker's range of possible attacks, which will be provided in the next sections.
	\subsubsection{Dynamic Controller}
	Consider the following distributed dynamic controller designed for the vehicle platoons:
	\begin{align}
		\label{eq:DynamicController}
		\begin{split}
			u_i&=e_iK\sum_{j \in \mathcal{N}_i}(x_i-x_j+d_{ij})\\&+e_isgn\Big(K\sum_{j \in \mathcal{N}_i}x_i-x_j+d_{ij}\Big)
		\end{split}
	\end{align}
	where $e_i$ is a time-varying coupling gain associated with the $i_{th}$ follower and it is governed by the following dynamics:
	\begin{align*}
		\begin{split}
			\dot{e}_i&=\tau \Big[\sum_{j \in \mathcal{N}_i}x_i-x_j+d_{ij}\Big]^T\Gamma \Big[\sum_{j \in \mathcal{N}_i}x_i-x_j+d_{ij}\Big]\\&+\tau ||K\sum_{j \in \mathcal{N}_i}x_i-x_j+d_{ij}||_1
		\end{split}
	\end{align*}
	where $\Gamma$ is a constant gain matrix, $K$ the gain defined later in Lemma 2, $e_i$ is a time-varying coupling gain associated with the $i_{th}$ follower, and $sgn(.)$ is the signum function defined component-wise. The proof of the following Lemma is provided in Appendix A.
	
	\textbf{Lemma 1}: Given Assumptions 1 and 2, the followers will track the leader in \textbf{attack free scenarios} under the control protocol (\ref{eq:DynamicController}) with $K=-B^T P^{-1}$ and $\Gamma=P^{-1}BB^TP^{-1}$, where $P>0$ is the solution to the following Ricatti inequality:
	\begin{align}
		\label{eq:Ricatti}
		AP+PA^T-2BB^T <0
	\end{align}
	Moreover, the coupling gains $e_i$ will converge to some finite steady-state value.

	\subsubsection{Static Controller}
	Consider the following control protocol for the follower agents $i=1,...,N$:
	\begin{align}
		\label{eq:StaticController}
		\begin{split}
			u_i&=c_1 K \sum_{j \in \mathcal{N}_i}(x_i-x_{j}^{c}+d_{ij}) \\&+ c_2sgn(K\sum_{j \in \mathcal{N}_i}(x_i-x_{j}^{c}+d_{ij}))
		\end{split}
	\end{align}
	where $c_1>0$, $c_2>0$ are constant coupling gains and $K$ is the gain matrix. Note that the following Lemma is an extension of theorem 1 in \cite{li2012distributed} and is therefore omitted for the sake of brevity.
	
	\textbf{Lemma 2}: Given Assumptions 1 and 2, under the distributed tracking controller (\ref{eq:StaticController}), the follower agents will track the leader in \textbf{attack free scenarios} with $c_1>\frac{1}{\lambda_1}$, $c_2>\gamma$, and $K=-B^TP^{-1}$, where $\lambda_1,...,\lambda_N$ are the eigenvalues of $\mathcal{L}_1$ and $P>0$ is a solution to the following well-known inequality:
	\begin{align}
		AP+PA^T-2BB^T <0
	\end{align}
	
	\textbf{Remark 1}: It should be noted that the static controller (\ref{eq:StaticController}) is not fully distributed since the vehicles need to know some information about the Laplacian matrix (since $c_1$ is chosen based on $\lambda_1$). Furthermore, the follower agents also need to know the upper bound on the leader's input (since the design of $c_2$ is dependent on $\gamma$). However, these requirements were not needed in the distributed dynamic controller (\ref{eq:DynamicController}) provided in previous section. 
	
	\textbf{Remark 2}: It can be observed that the dynamic (\ref{eq:DynamicController}) and static (\ref{eq:StaticController}) control protocols are more complex in comparison with the control protocols in \cite{han2024resilient} and \cite{he2025resilient}. This is due to the fact that in these papers, the leader's input $u_0$ is assumed to be zero (i.e., $u_0=0$) while in our work, we have considered a time-varying input signal for the leader and we have assumed that the follower vehicles do not have access to the leader's input signal.
	
		\subsection{Cyber-Attack Model}
	Since the Vehicle-to-Vehicle (V2V) communication channels are vulnerable to attacks, the attacker is capable of launching False Data Injection (FDI) attacks on the communication channels from vehicle $j$ to vehicle $i$, i.e.,:
	\begin{align}
		\label{eq:CyberAttackDefinition}
		x_{j}^{c}=x_j+a_{ji}
	\end{align}
	where $a_{ji}$ is the attack signal injected to the state information sent from vehicle $j$ to vehicle $i$ and $x_{j}^{c}$ is the corrupted state information received at vehicle $i$ from vehicle $j$.
	
	By extending the residuals in \cite{taheri2020undetectable} and \cite{zhang2024ripple} to vehicle platoons, we consider the following residual in each vehicle to detect anomalies:
	\begin{align}
		\label{eq:residual}
		r_{x}^{i}=\sum_{j \in \mathcal{N}_i}||x_{i}-x_{j}^{c}+d_{ij}||
	\end{align}
	where $d_{ij}=[(i-j)d_0,0,0]$ and $d_{i0}=d_i$.
	
	\textbf{Definition 2}: The attacks on the communication links are stealthy if in the presence of attacks, (i.e., $a_{ji}\ne 0, \forall t>0$), the following equation is satisfied for the $r_{x}^{i}$:
	\begin{align}
		\label{eq:stealthiness}
		\lim_{t\to \infty}r_{x}^{i}=0, \ for \; all \; i=1,...,N
	\end{align}
	
	\textbf{Remark 3}: Note that the residual signal (\ref{eq:residual}) will generate false alarms in the transient time of the platoon in normal, cyber-attack free scenarios. This demonstrates the need to develop better residual generation mechanisms that handle both transient and steady state phases of the system response. Further discussions are provided in Section \ref{sec:Discussions}.
	
	\textbf{Definition 3}: The attacks on the communication channels have the following two objectives:
	\begin{enumerate}
		\item Remain stealthy based on equation (\ref{eq:stealthiness}) in Definition 2.
		\item Lead the under-attacked subset of the vehicles to track the attackers state $x_a$ while satisfying the required spacing. The attackers state is governed by the following dynamics:
		\begin{align}
			\label{eq:AttackerDynamics}
			\dot{x}_a=Ax_a+Bu_a
		\end{align}
		where $u_a$ is the attackers reference signal and it is bounded, i.e., $||u_a||_{\infty} \le \gamma_2$ with $\gamma_2$ a positive constant.
	\end{enumerate}
	\subsection{Objectives}
	In this paper, we follow two main objectives. First, we aim to design attack signals $a_{ji}$ to ensure that the attacker can achieve its goals as defined in Definition 3. In this first objective, we will investigate the possibility and conditions of achieving these goals in different scenarios such as the case where the attacker can launch attacks on all the communication channels and the case where the attacker can only launch attacks on the communication channels between the follower vehicles.
	
	Second, we want to investigate the effects of the systems control protocol, i.e. how the design of the control input signal $u_i$ influences the design of the attack signal $a_{ji}$. We investigate the distributed static (\ref{eq:StaticController}) and dynamic (\ref{eq:DynamicController}) control protocols for the vehicles to see how the attacker needs to adapt to different control protocols to achieve its objectives as defined in Definition 3.

	\section{Cyber-Attack Design with Dynamic Controller}
	\label{sec:DynamicController}

	In this section, we consider that the vehicles in the platoon are utilizing the dynamic controller (\ref{eq:DynamicController}). Two scenarios for the attacker will be investigated in this section:
	\begin{enumerate}
		\item The attacker can inject FDI attacks on all the communication channels, including the outgoing communication channels of the leader and the communication channels between the followers.
		\item While the outgoing communication channels of the leader are attack free, the attacker can launch attacks on the communication channels between the follower vehicles.
	\end{enumerate}
	The reason behind the above categorization is due to the fact that in some cases, the leader vehicle has secure communication capabilities (e.g., refer to \cite{adoni2024intelligent}) while the follower vehicles do not posses such capabilities. Therefore, we aim to analyze the possibility of achieving the attackers objectives in these two scenarios.
	
	Furthermore, note that due to the fact that we have considered attacks on the communication channels, the leader vehicle in the considered platoon topology, \textit{k-Nearest Neighbor Leader Tracking}, cannot be subject to attacks itself due to the fact that the leader vehicle does not have any incoming communication channels.
	\subsection{All Communication Channels are Vulnerable to Attacks}
	\begin{figure}
		\begin{center}
			\includegraphics[width=8.4cm]{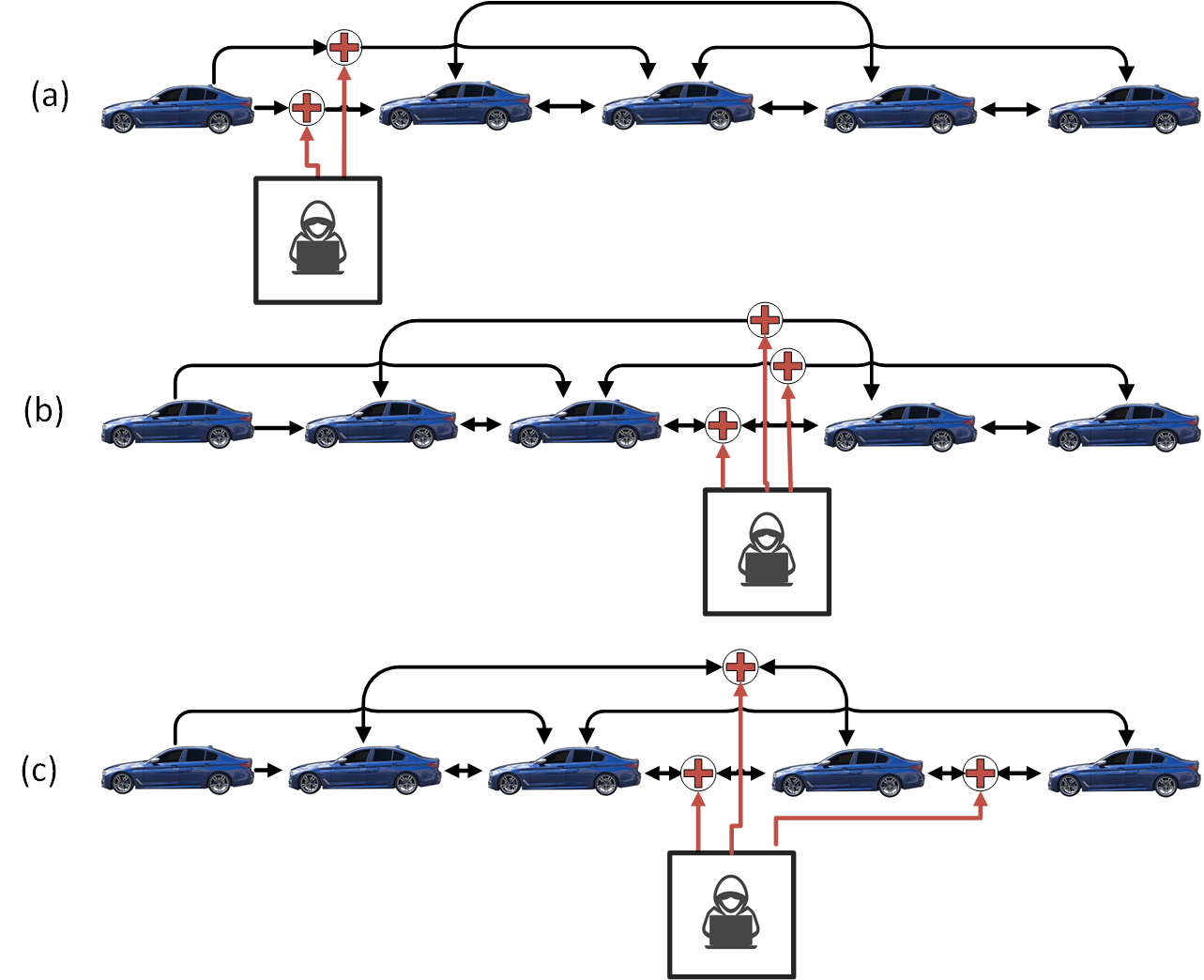}    
			\caption{Cyber-attack scenarios based on the attackers resources and objectives in the \textit{k-Nearest Neighbor Leader Tracking} topology: (a) attacks on the communication channels of the leader to lead all the follower vehicles to track the attacker, (b) attacks on the followers not directly connected to the leader to lead the follower vehicles that do not have any direct communication channels from the leader to track the attacker, (c) attacks on a subset of followers not directly connected to the leader to lead this subset to track the attacker.}
			\label{fig:PlatoonUnderAttack}
		\end{center}
	\end{figure}
	In this section, we consider that all of the communication channels, including the communication channels of the leader can be subject to attacks with an example provided in Fig. \ref{fig:PlatoonUnderAttack}(a). Therefore, the following attack signal is proposed:
	\begin{align}
		\label{eq:CyberAttackDynamic1}
		a_{0i}=x_a-x_0
	\end{align}
	where $i \in \mathcal{V}_{d}$. The proof of the following theorem is provided in Appendix \ref{appendix:2}.
	
	\textbf{Theorem 1}: Under Assumptions 1 and 2, the attack signal (\ref{eq:CyberAttackDynamic1}) will remain stealthy while causing the follower vehicles to track the attackers state $x_a$ with the required spacing.
	
	We now investigate the required resources in order for the attackers to launch the attack signal (\ref{eq:CyberAttackDynamic1}). The attacker's resources can be broadly categorized into the following three categories \cite{teixeira2012attack}:
	\begin{enumerate}
		\item \textbf{Disclosure Resources}: Ability to observe or eavesdrop the system signals (e.g., sensor data, control commands, communication packets) without altering them.
		\item \textbf{Disruption Resources}: Ability to modify, block, or inject signals into the system (e.g., false data injection and jamming).
		\item \textbf{System Knowledge}: \textit{a priori} knowledge of the system model, dynamics, or topology, allowing the attacker to design stealthy or targeted attacks.
	\end{enumerate}
	
	Therefore, in order for the attacker to launch the attack signal (\ref{eq:CyberAttackDynamic1}), the attacker requires the following resources:
	
	\textit{Disruption Resources}: The attacker needs to launch attacks on all the communication channels of the leader. 
	
	\textit{Disclosure Resources}: The attacker needs to obtain the state of the leader $x_0$.
	
	\textit{System Knowledge}: The attacker needs some graph information to know which communication links originate from the leader. However, knowledge about the system dynamics is not required.

	\textbf{Remark 4}: It is worth noting that if the attacker has access to both disclosure and disruption resources, the attack signal defined in (\ref{eq:CyberAttackDynamic1}) is a viable and practical choice, as it does not require any knowledge of the underlying system dynamics. As established in Theorem 1, such an attack signal is sufficient to drive the followers to track the attacker's state while satisfying the required spacing between the vehicles, even in the absence of model information.
	
	\subsection{No Attacks on the Communication Channels of the Leader}
	In this section, we consider the case where the communication links of the leader are attack free. Consider the following cases corresponding to the attackers second objective based on Definition 3:
	\begin{enumerate}
		\item The attackers want all the follower vehicles that are not connected directly to the leader to follow their state.
		\item The attackers want a subset of the follower vehicles that are not connected directly to the leader to follow their state.
		\item The attackers want all the follower vehicles to follow their state.
	\end{enumerate}
	In this section, we will show that the attackers cannot achieve both of their objectives for the follower vehicles that are receiving information directly from the leader (denoted by the set $\mathcal{V}_d$). However, the attackers can achieve both of their objectives for the follower vehicles that are not receiving information directly from the leader (denoted as the set $\mathcal{V}_{nd}$).
	\subsubsection{The Follower Vehicles Not Directly Connected To the Leader}
	The following theorem is provided to demonstrate how the attacker can lead the vehicles in the set $\mathcal{V}_{nd}$ to track the attacker while remaining stealthy and its proof is provided in Appendix \ref{appendix:3}.
	
	\textbf{Theorem 2}: The vehicles in the set $\mathcal{V}_{nd}$ will track the attacker while remaining stealthy if the following conditions are satisfied:
	\begin{itemize}
		\item All the communication channels going from the set $\mathcal{V}_{d}$ to the set $\mathcal{V}_{nd}$ should be subject to the following attack:
		\begin{align}
			\label{eq:cyber-attackdynamic2}
			a_{ij}=x_a - x_0+d_i
		\end{align}
		where $i \in \mathcal{V}_{d}$ and $j \in \mathcal{V}_{nd}$.
		\item All the communication channels going from the set $\mathcal{V}_{nd}$ to the set $\mathcal{V}_{d}$ should be subject to the follower attack:
		\begin{align}
			\label{eq:cyber-attackdynamicstealthy}
			a_{ji}=x_0-d_j-x_a+d_{j}^{c}
		\end{align}
		where $d_{j}^{c}=[j^cd_0,0,0]$ where \( j^{c} \in \{1, \dots, |\mathcal{V}_{nd}|\} \) is defined as the local index of vehicle \( j^{c} \) within the ordered subset \( \mathcal{V}_{nd}\).
	\end{itemize}

	The following resources are necessary for the attacker to achieve its objectives:
		
		\textit{Disruption Resources}: The attacker needs to launch attacks on all the incoming and outgoing communication channels of the vehicles in the set $\mathcal{V}_{nd}$. 

		\textit{Disclosure Resources}: The attacker needs to obtain the state of the leader $x_0$.
		
		\textit{System Knowledge}: The attacker needs graph information to know some the communication links to and from the set $\mathcal{V}_{nd}$ and the index of the vehicles in the platoon (to obtain $d_i$ and $d_j$). However, knowledge about the system dynamics is not required.

	\textbf{Remark 5}: One of the benefits of the proposed attacks (\ref{eq:cyber-attackdynamic2}) and (\ref{eq:cyber-attackdynamicstealthy}) is that different from \cite{taheri2020undetectable} (where the online outputs and state estimates of each under-attacked vehicle is required) and \cite{tsang2020sparse} (where the online states estimates of all the vehicles are required), the designed attacks would only require the online state information of the leader for its disclosure resources. This is especially important in FDI attacks in the physical layer (e.g., Symbol Flipping \cite{popper2011investigation}) where the attacker transmits a signal to be combined with the original signal of the system, modifying the values of the received signal.
	
	\textbf{Remark 6}: Note that while the attacker will be detected by the residual (\ref{eq:residual}) in the follower vehicles in the set $\mathcal{V}_{d}$, the detection happens only if the attack happens in the steady state. However, if the attacker starts the attack during the transient time, the vehicles in the set $\mathcal{V}_d$ cannot distinguish presence of cyber-attacks from the false alarms of the residual (\ref{eq:residual}) in the transient time. This fact will be further illustrated in Section \ref{sec:Numerical Example}.
	
	\subsubsection{A Subset of The Follower Vehicles Not Directly Connected To the Leader}
	We now turn our focus to when the attacker wants to ensure that a subset of the set ${\mathcal{V}_{nd}}$ tracks the attacker. Let us denote this subset as ${\mathcal{V}_{nd}^{a}}$. Furthermore, denote $\mathcal{V}_{nd}^{c}$ as the nodes of the set $\mathcal{V}_{nd}$ that are still connected to the leader (i.e., there exists a attack free path from the leader to the vehicles in this set) and $\mathcal{V}_{nd}^{dc}$ as the nodes of the set $\mathcal{V}_{nd}$ that are disconnected from the leader (i.e., there does not exist a attack free path from the leader to the vehicles in this set). Therefore,
	to achieve the attackers objectives, we have the following theorem with its proof provided in Appendix \ref{Appendix:4}.

		\textbf{Theorem 3}: Let $V_{nd}^a \subseteq V_{nd}$ denote a subset of follower vehicles that the attacker aims to manipulate.  The vehicles in $V_{nd}^a$ will track the attacker while remaining stealthy if the attacker injects the following attack signals:
		\begin{enumerate}
			\item For communication links between the set $\mathcal{V}_{nd}^{a}$ and the set $\mathcal{V}_{nd}^{c}$ (the vehicles that are connected to the leader):
			\begin{align}
			\label{eq:Cyber-attack on a subset of followers}
			a_{ij} &= x_a - x_0+d_i\\
			\label{eq:Cyber-attack on a subset of followers stealthy}
			 a_{ji} &= x_0 - x_a-d_j+d_{j}^{c}
			\end{align}
			where $i$ belongs to the set $\mathcal{V}_{nd}^{c}$ that have a directed path from the leader, $j$ belongs to the under-attacked subset of the vehicles $\mathcal{V}_{nd}^{a}$.
			\item For communication links between the set $\mathcal{V}_{nd}^{a}$ and the set $\mathcal{V}_{nd}^{dc}$ (the vehicles that are disconnected from the leader):
			\begin{align}
				\label{eq:Cyber-attack on a subset of followers not connected}
				a_{ij} &= x_a-x_i-d_{j}^{c}-d_{ij}\\
				\label{eq:Cyber-attack on a subset of followers not connected stealthy}
				 a_{ji} &= x_i - x_a-d_{ji}+d_{j}^{c}
			\end{align}
			where $i$ belongs to the set $\mathcal{V}_{nd}^{dc}$ and $j$ belongs to the under-attacked subset of the vehicles $\mathcal{V}_{nd}^{a}$.
		\end{enumerate}
	
	The following resources are necessary for the attacker to achieve its objectives:
	
 		\textit{Disruption Resources}: The attacker needs to launch attacks on all the incoming and outgoing communication channels of the vehicles in the set $\mathcal{V}_{nd}^{a}$. 
 		
		 \textit{Disclosure Resources}: The attacker needs to obtain the state of the leader $x_0$, along with the state of the vehicles in the set $\mathcal{V}_{nd}^{dc}$ that are disconnected from the leader.
		 
		 \textit{System Knowledge}: The attacker needs the entire topology information to know the indices of the vehicles and whether a specific vehicle is connected to the leader or not . However, knowledge about the system dynamics is not required.
	
	\textbf{Remark 7}: It should be noted that the attackers cannot achieve both of their objectives for all of the followers if the communication channels of the leader are secure. One can easily observe that in this scenario, since the vehicle is directly receiving information from the leader, any deviation from the tracking of the leader will be detected by the residual (\ref{eq:residual}). Therefore, while it is possible to degrade the following vehicles, the attackers cannot achieve both of their objectives in this case.

	In the next section, we will investigate the possibility of achieving both of the attacker's objectives when the vehicles are utilizing another control protocol, namely a distributed static controller.

	\section{Cyber-Attack Design with Static Controller}
	\label{sec:StaticController}
	Here, we consider the static controller (\ref{eq:StaticController}) for the vehicles and that all of the communication channels, including the communication channels of the leader, are vulnerable to attacks. The following attacks are design for the outgoing communication channels of the leader:
	\begin{align}
		\label{eq:AttackDesignStatic}
		a_{0i}=x_a-x_0
	\end{align}
	Then, we have the following theorem with its proof provided in Appendix \ref{Appendix:5}.
	 
	\textbf{Theorem 4}: Under Assumptions 1 and 2, the attack signal (\ref{eq:AttackDesignStatic}) will remain stealthy while causing the follower agents to track the attacker's state if $\gamma_2 \le \gamma$, i.e., the upper bound of the attacker's reference signal $u_a$ is less than or equal than the upper bound of the leader's input $u_0$.

	It can be seen that the attacker requires the following resources for this case:
	
	\textit{Disruption Resources}: The attacker needs to launch attacks on all the communication channels of the leader. 
	
	\textit{Disclosure Resources}: The attacker needs to obtain the state of the leader $x_0$.
	
	\textit{System Knowledge}: The attacker needs some graph information to know which communication links originate from the leader and what is the value of $\gamma$ to ensure $\gamma_2 < \gamma$. However, knowledge about the system dynamics is not required.
	
	Note that the attack signal (\ref{eq:AttackDesignStatic}) was required to satisfy $u_a \le u_0$ and in order for the attacker to cause the platoon to follow the attacker while the attacker's reference signal satisfies $r_{a}\ge \gamma$, they can choose the attack signal as:
	\begin{align}
		\label{eq:cyber-attacksignalstatic2}
		a_{0i}=x_a-x_0 - \frac{c_3-c_2}{c_1 ||K||}sgn(K\sum_{j\in \mathcal{N}_i }(x_i-x_j+d_{ij}))
	\end{align}
	where $c_3$ is a positive constant satisfying $ \gamma_2 \le c_3$. 
	
	Next, we show that the attack signal (\ref{eq:cyber-attacksignalstatic2}) will cause all the agents to follow the attacker's reference signal. However, this attack will not be stealthy.
	
	Given the attack signal (\ref{eq:cyber-attacksignalstatic2}), we will obtain the following for the agents that are directly connected to the leader (the set $\mathcal{V}_d$):
	\begin{align}
		\begin{split}
			\dot{x}_i &= Ax_i + BKc_1 \Bigg[ 
			\sum_{j \in \mathcal{N}_i} (x_i - x_j+d_{ij})
			- x_a + x_0  \\&+ \frac{c_3 - c_2}{c_1 \|K\|} \, sgn(K\sum_{j \in \mathcal{N}_i}(x_i-x_j+d_{ij}))
			\Bigg] \\ &+ Bc_2sgn\Big(K\sum_{j\in \mathcal{N}_i }(x_i-x_j+d_{ij})-x_a+x_0\\&+\frac{c_3 - c_2}{c_1 \|K\|} \, sgn(K\sum_{j \in \mathcal{N}_i}(x_i-x_j+d_{ij}))   \Big)
		\end{split}
	\end{align}
	Given that $sgn(a)=sgn(a+bsgn(a))$ where $b>0$ is a constant, we obtain:
	\begin{align}
		\begin{split}
			\dot{x}_i &= Ax_i + BKc_1 \Bigg[ 
			\sum_{j \in \mathcal{N}_i} (x_i - x_j+d_{ij})
			\Bigg] \\&+  Bc_3sgn\Big(K\sum_{j\in \mathcal{N}_i }(x_i-x_j+d_{ij}) \Big)
		\end{split}
	\end{align}
	Then, we have the following theorem with its proof provided in Appendix \ref{Appendix:6}.
	
	\textbf{Theorem 5}: The attack signal (\ref{eq:cyber-attacksignalstatic2}) will cause all the follower vehicles to follow the attacker. However, it will not be stealthy.
	
	It can be seen that for this case, the attacker requires the following resources:
	
	\textit{Disruption Resources}: The attacker needs to launch attacks on all the communication channels of the leader. 
	
	\textit{Disclosure Resources}: The attacker needs to obtain the state of the leader $x_0$ while also obtaining the states of the neighboring vehicles of the set $\mathcal{V}_d$.
	
	\textit{System Knowledge}: The attacker needs some graph information to know which communication links originate from the leader, knowledge about the control gains (to obtain $K$, $c_1$, and $c_2$), and the overall control protocol structure (to design the term $\frac{c_3-c_2}{c_1 ||K||}sgn(K\sum_{j\in \mathcal{N}_i }(x_i-x_j+d_{ij}))$).
	
	Note that scenarios in which the leader’s communication channels are secure under static controllers are analogous to those when dynamic controllers are employed. One can easily extend the results of Theorem 2 and Theorem 3 to the static case and therefore, they are not provided here for the sake of brevity.
	
		\section{Discussions \& Implications for Resiliency of Vehicle Platoons}
	\label{sec:Discussions}
	\begin{table*}[t]
		\centering
		\caption{Summary of attack scenarios, required resources, and stealthiness}
		\label{tab:attack_summary}
		\renewcommand{\arraystretch}{1.3}	
		\begin{tabular}{|p{2.3cm}|c|c|p{2.7cm}|p{2.7cm}|p{2.5cm}|p{1cm}|}
			\hline 
			\textbf{Scenario / Affected vehicles} &\textbf{Controller}&\textbf{Leader Links}& \textbf{Disruption Resources} & \textbf{Disclosure Resources} & \textbf{System Knowledge} & \textbf{Stealthy?} \\
			\hline
			\textbf{Theorem 1} -- All followers 
			&Dynamic
			& Vulnerable
			& Inject FDI on all leader links 
			& Leader state $x_0$ 
			& Leader’s outgoing links 
			& Yes \\
			\hline
			\textbf{Theorem 2} -- Followers in set $\mathcal{V}_{nd}$
			& Dynamic
			& Secure
			& Inject FDI on all in/out links of $V_{nd}$ 
			& Leader state $x_0$ 
			& links to/from $V_{nd}$, follower indices 
			& Yes \\
			\hline
			\textbf{Theorem 3} -- Subset of $V_{nd}$ vehicles
			& Dynamic
			& Secure
			& Inject FDI on all in/out links of $V_{nd}^{a}$ 
			& Leader state $x_0$ and states of disconnected followers $V_{nd}^d$ 
			& Full topology knowledge 
			& Yes  \\
			\hline
			\textbf{Theorem 4} -- All followers with $u_a < \gamma$
			&Static
			& Vulnerable
			& Inject FDI on all leader links 
			& Leader state $x_0$ 
			& Leader's outgoing links + leader input bound $\gamma$ 
			& Yes 
			\\
			\hline
			\textbf{Theorem 5} -- All followers with $u_a > \gamma$
			& Static
			& Vulnerable
			& Inject FDI on all leader links 
			& Leader + neighbor states of $V_d$ 
			& Leader's outgoing links, control protocol and gains
			& No\\
			\hline
		\end{tabular}
	\end{table*}

	The attack scenarios summarized in Table~\ref{tab:attack_summary} highlight
	that stealthy attacks depend heavily on the
	attacker’s available resources. By aligning resiliency measures to the
	specific exploited resources in each scenario, we can design more effective
	and resource-aware defenses and mitigation strategies. In the following sections, we provide several resiliency and mitigation approaches against stealthy attacks.

	\subsection{Limiting Disclosure Resources}
	In different scenarios, access to the leader’s state $x_0$ or certain follower states	is a critical disclosure resource in order for the attacker to launch attacks on the vehicle platoon. Preventing such access can significantly
	restrict the attacker’s ability to design stealthy attacks. Different privacy techniques has been investigated in the literature that aim to limit the attackers access to vehicle states:
	\begin{itemize}
		\item \textit{Perturbation-based methods:}\cite{wan2023differentially} intentionally perturb or mask transmitted states with bounded noise to conceal exact information.
		\item \textit{Encryption-based methods:}\cite{sultangazin2020symmetries} use homomorphic encryption or
		output masking to mask the transmitted data.
		\item \textit{Observability-based methods:}\cite{zhang2022privacy} redesign controllers and system structure to reduce observability of states for potential eavesdroppers.
	\end{itemize}
	While these approaches target the disclosure dependency evident in
	Table~\ref{tab:attack_summary}, designing an integrated privacy preserving control protocol and detection mechanism is an avenue of research for our future studies.
	
	\subsection{Switching Graph Topologies}
\begin{figure}
		\begin{center}
			\includegraphics[width=8.4cm]{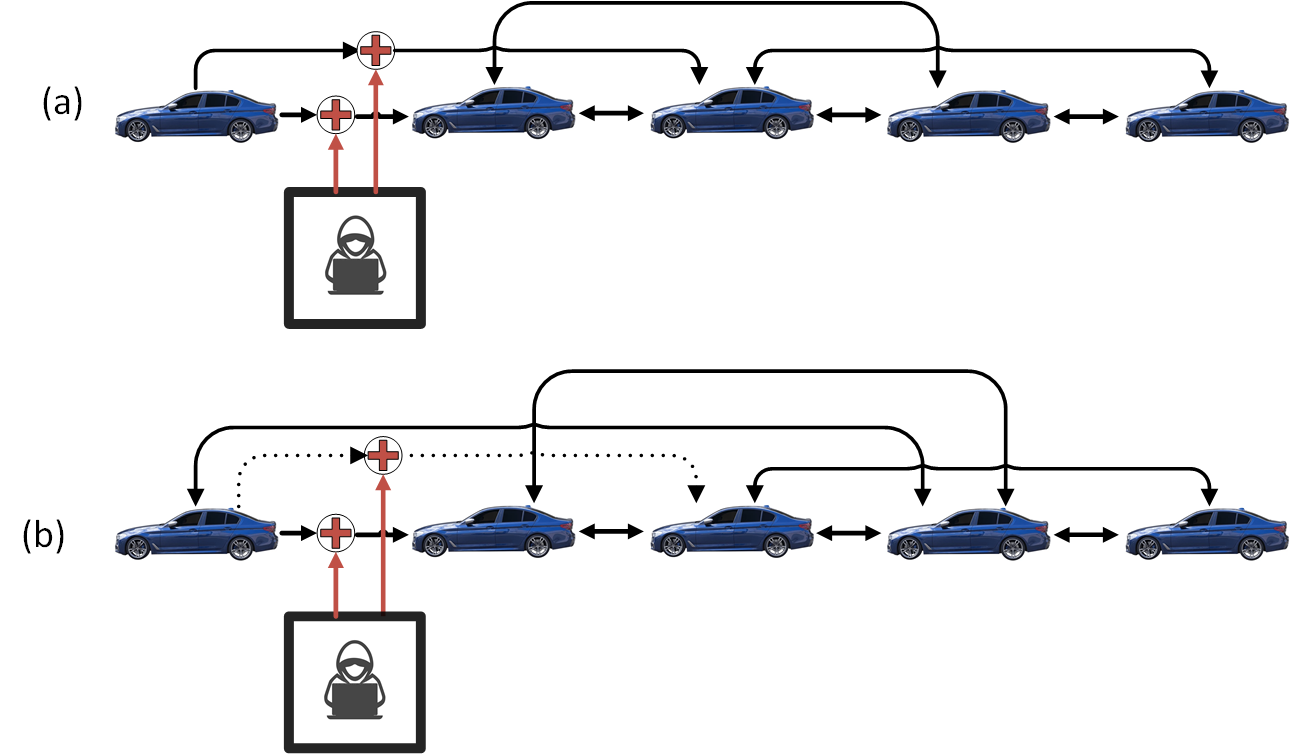}    
			\caption{Switching the communication channels and the topology in order to increase resiliency against attacks. (a) The system where the communication channels of the leader are subject to attacks. (b) The leader disconnects from the second follower and connects to the third follower, sending information to the third follower without being corrupted by the attacks. }
			\label{fig:PlatoonsUnderAttackSwitching}
		\end{center}
	\end{figure}

	As shown in Table \ref{tab:attack_summary}, attackers require different level of knowledge of the
	communication topology. Furthermore, they would also need to be able to launch cyber-attacks on the communication channels. To counter these:
	\begin{itemize}
		\item \textit{Switching topologies:} dynamically alter communication
		links, increasing the attacker’s effort to maintain accurate topology
		knowledge and creating opportunities for detection (as demonstrated in Fig. \ref{fig:PlatoonsUnderAttackSwitching}). Another approach is to switch between different platoon topologies. For example, switching from Nearest Neighbor topology to Leader-to-All topology.
		\item \textit{Random link activation/deactivation:} stochastically enable or disable communication channels according to predefined probabilities, making it impossible for attackers to reliably predict future connectivity and thereby forcing them to operate under persistent uncertainty.
		
	\end{itemize}
	This class of defenses specifically targets disruption and graph knowledge
	resources required in Theorems~1--5.
	
	\subsection{Improving Detection Mechanisms}
	Stealthiness in Theorems~1--4 arises because residual~(12) is insensitive to
	attacks once the system has reached steady state. This limitation allows
	attackers to remain undetected during transients. To address this, detection
	mechanisms must be extended to:
	\begin{itemize}
		\item Distinguish genuine transients from malicious attacks,
		thereby avoiding false alarms during transient time.
		\item Integrate distributed estimation of the neighboring agents with the detection mechanism
		\item Incorporate moving-target approaches where residual generation
		itself varies dynamically, limiting the attacker’s ability design attacks based on the residual and	remain stealthy.
	\end{itemize}
	By utilizing the above approaches, one can prevent stealthiness of attacks, which paves the way for recovery mechanisms to be activated.
	
	\subsection{Imposing Constraints}
	Imposing constraints can reduce the attacker’s feasible set of possible attacks that achieves both of the attackers objectives (per Definition 3). For example:
	\begin{itemize}
		\item Constraining the leader’s input bound $\gamma$ (Theorem~4) makes
		stealthy attacks feasible only within narrow limits.
		\item Increasing the communication links to increase graph robustness and utilizing MSR approaches (as stated in Section \ref{subsec:Related Work}).
	\end{itemize}
	Note that these approaches can be combined together as well to increase redundancies in the system.

	\section{Numerical Case Study}
	\label{sec:Numerical Example}
In this section, we present numerical simulations of a vehicle platoon composed of one leader and four followers under various conditions, including attacks on the leaders outgoing links, attack on the set $\mathcal{V}_{nd}$, selective attacks on a subset of followers, and attack on the leaders outgoing links with the static controller.

\begin{figure}
	\begin{center}
		\includegraphics[width=8.2cm]{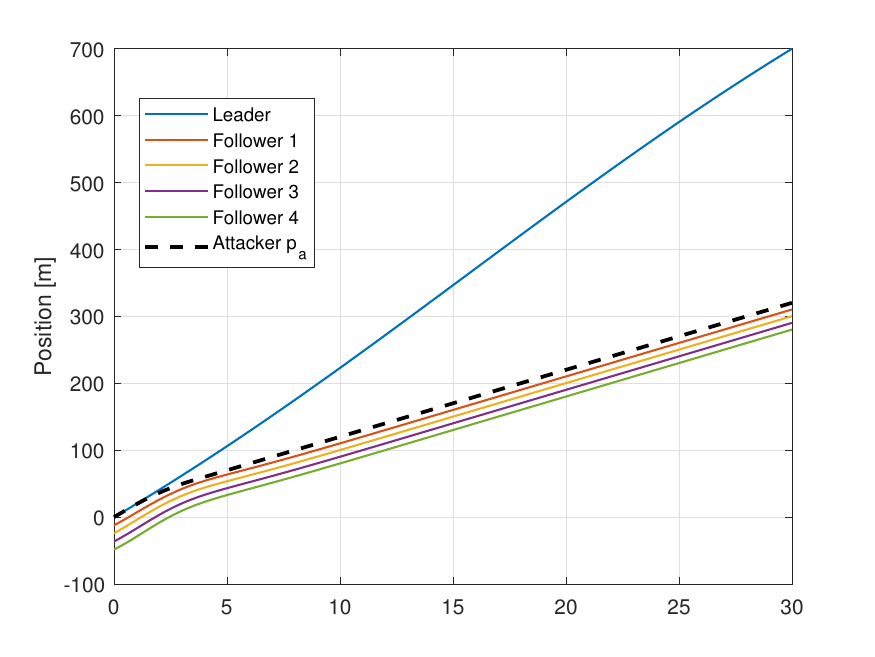}    
		\caption{First scenario where the communication channels of the leader are compromised.}
		\label{fig:FromStartPositionAll}
	\end{center}
\end{figure}

\begin{figure}
	\begin{center}
		\includegraphics[width=8.4cm]{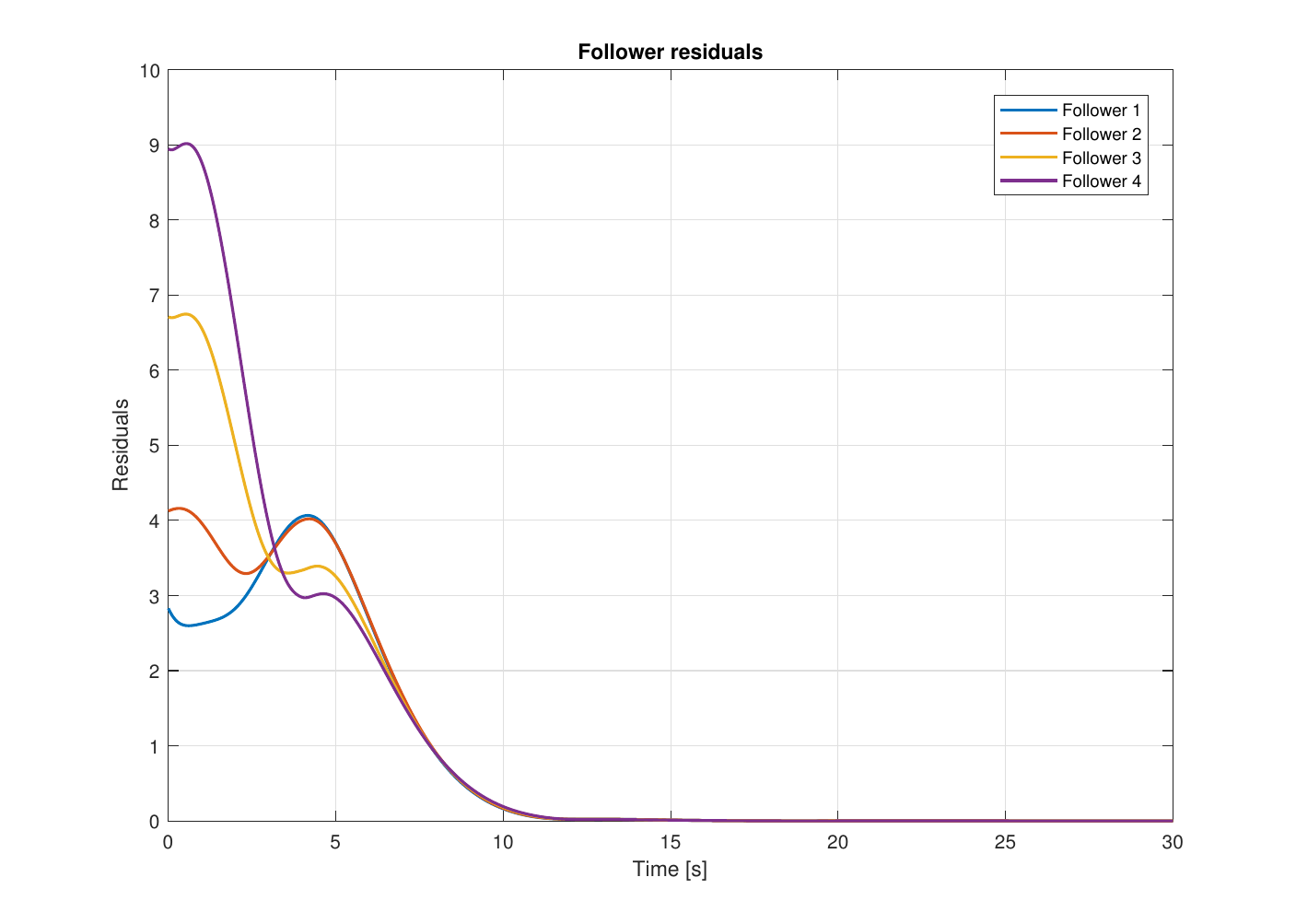}    
		\caption{Residuals of the followers where the communication channels of the leader are compromised.}
		\label{fig:FollowerResidualsAllUnderAttack}
	\end{center}
\end{figure}

We consider a platoon of $N=5$ vehicles, where the first vehicle is the leader and the remaining vehicles are followers where $k=2$ in the \textit{k-Nearest Neighbor} topology. The lag constant is set to $\tau_i=0.4$, the desired inter-vehicle gap is considered to be $20$ meters, and $K=[-0.7,-1.2,-0.05]$. 
By considering the control input of the leader to be $u_0=0.4(0.005sin(0.1t)+0.0075cos(0.1t))$, The leader acceleration $a_0$ and velocity $v_0$ dynamics will become:$
	v_0(t) = 20 + 0.05 \sin(0.1 t)$ and $
	a_0(t)=0.005cos(0.1t)$
Furthermore, the attackers acceleration is the same as the leader, while its velocity is $v_a(t)=10+0.05\sin(0.1t)$.
In the first scenario, we consider that the attacker launches FDI attacks on all the communication channels of the leader. The effects of the attack can be seen in Fig. \ref{fig:FromStartPositionAll}. As can be seen in this figure, the followers started following the attacker when the attack starts. Furthermore, Fig. \ref{fig:FollowerResidualsAllUnderAttack} demonstrates the residuals (\ref{eq:residual}) which demonstrates that the attack signal is stealthy. As mentioned in Section \ref{sec:Discussions}, a better residual is required in order to detect attacks when they start in the transient time.

\begin{figure}
	\begin{center}
		\includegraphics[width=8.2cm]{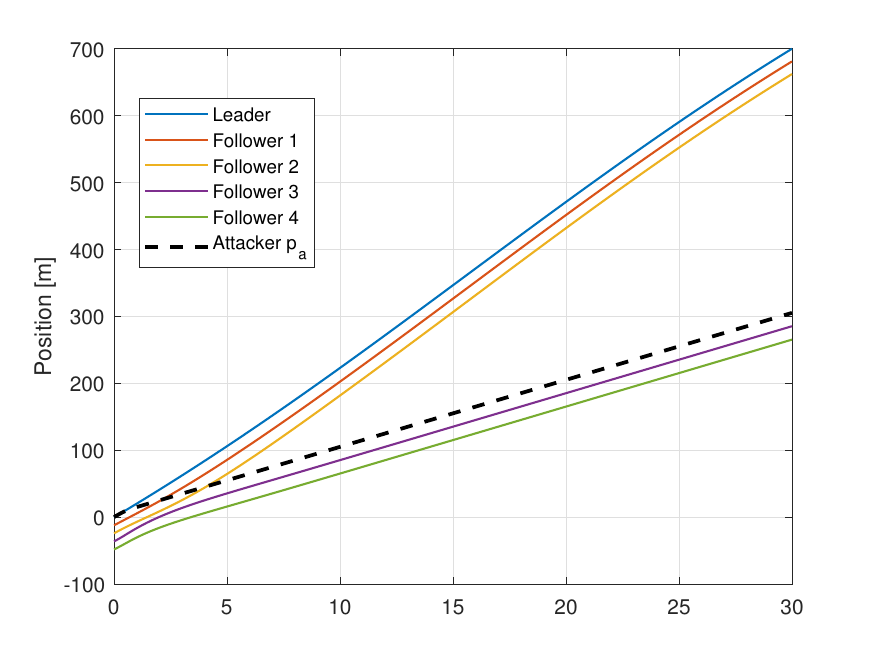}    
		\caption{Second Scenario where the attacker launches FDI attacks on the communication channels of follower vehicles $3$ and $4$.}
		\label{fig:FromStartPositionSomeSmooth}
	\end{center}
\end{figure}

In the second scenario, we consider that the communication channels of the leader are secure and the attacker is trying to make the follower vehicles that are not directly connected to the leader to follow its state trajectory. As can be seen in Fig. \ref{fig:FromStartPositionSomeSmooth}, by launching the attack signals (\ref{eq:cyber-attackdynamic2}) and (\ref{eq:cyber-attackdynamicstealthy}), the follower vehicles 3 and 4 are following the attacker. Note that the residuals are similar to the first scenario and is omitted for the sake of brevity.

\begin{figure}
	\begin{center}
		\includegraphics[width=8.2cm]{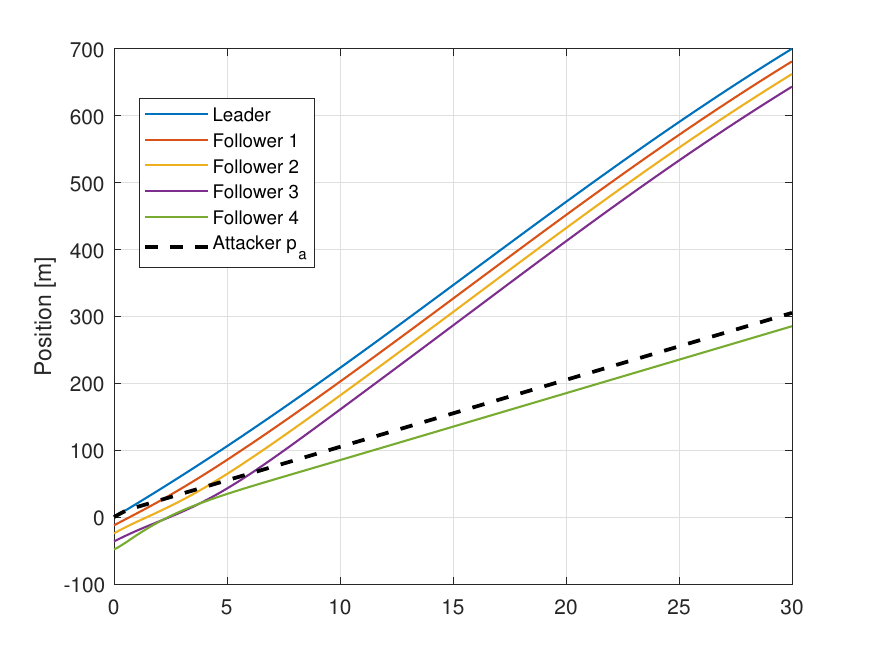}    
		\caption{Third Scenario where the attacker launches FDI attacks on the communication channels of the follower vehicle $4$.}
		\label{fig:FromStartPositionOneSmooth}
	\end{center}
\end{figure}

In the third scenario, we consider that the attacker attacks only follower 4 by launching the attack signals (\ref{eq:Cyber-attack on a subset of followers not connected}) and (\ref{eq:Cyber-attack on a subset of followers not connected stealthy}). As can be seen in Fig. \ref{fig:FromStartPositionOneSmooth}, Follower 4 follows the attacker while the rest of the vehicles follow the leader. Since the residual signals are similar to the first scenario, they are omitted for the sake of brevity.

\begin{figure}
	\begin{center}
		\includegraphics[width=8.2cm]{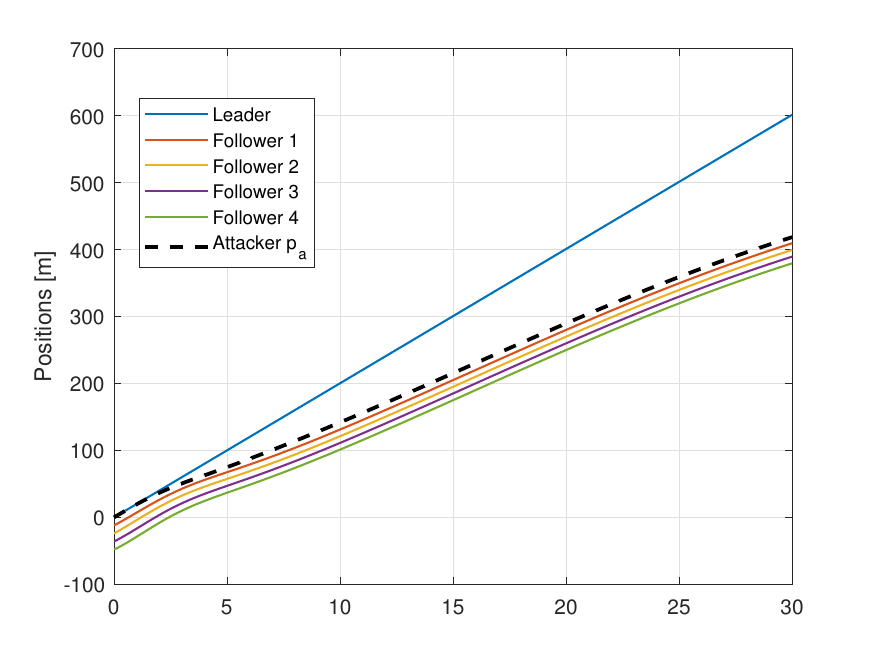}    
		\caption{Fourth Scenario where the attacker launches FDI attacks on the communication channels of the leader with static control protocol.}
		\label{fig:StatesStaticNonstealthy}
	\end{center}
\end{figure}

\begin{figure}
	\begin{center}
		\includegraphics[width=8.2cm]{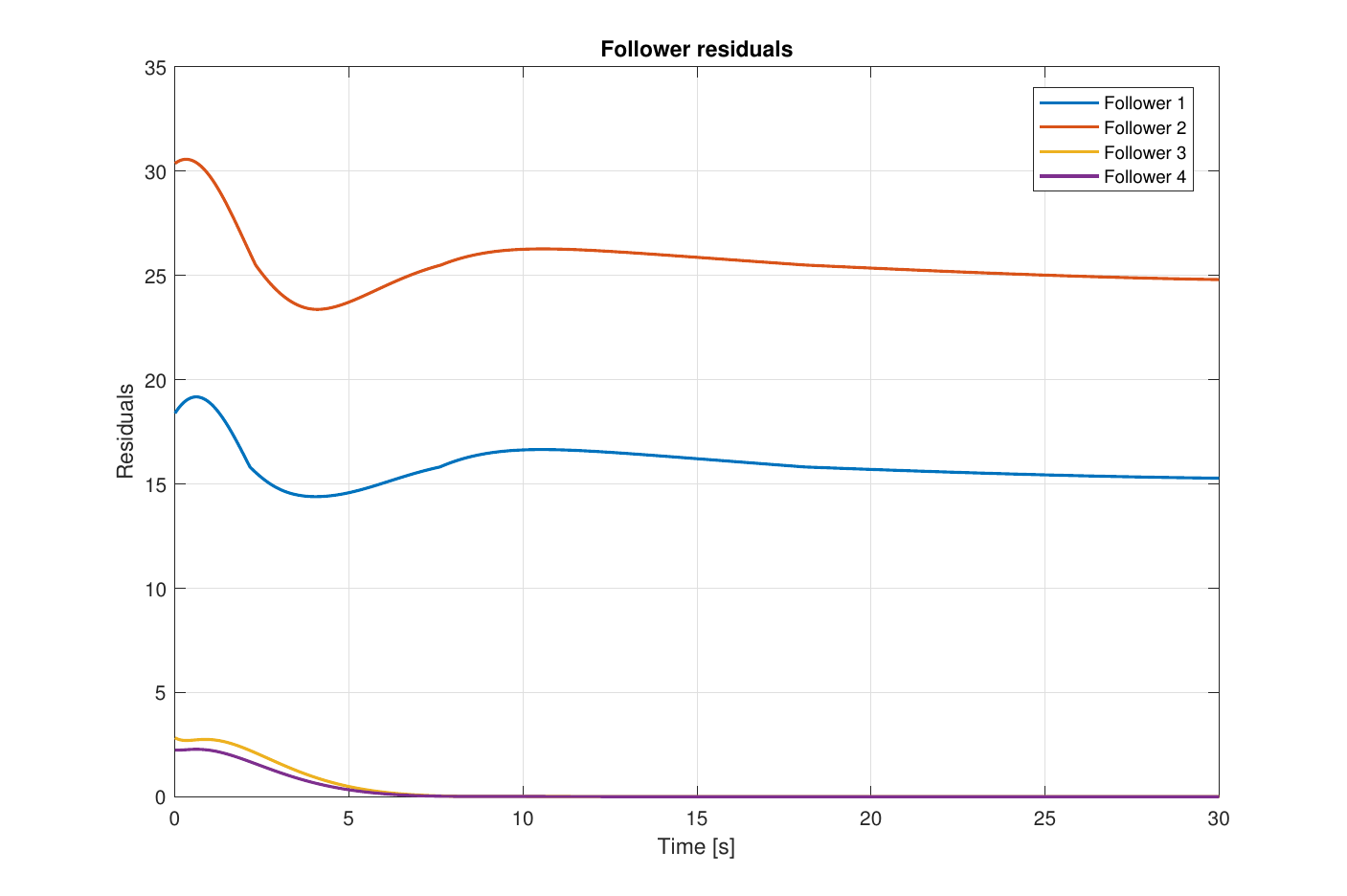}    
		\caption{Residuals in the fourth scenario where the attacker launches FDI attacks on the communication channels of the leader with static control protocol.}
		\label{fig:residualinStaticNonstealthy}
	\end{center}
\end{figure}

Finally, in the fourth scenario, we consider the static controller case with $c_1=2$, $c_2=1$ with the same $K$ as the dynamic controller case. If the attacker considers the same acceleration and velocity as the previous cases, the attack will be stealthy and similar to the dynamic controller case and due to this reason and for the sake of brevity, is not provided here. However, when the attacker has a velocity of $v_a(t)=10+5sin(0.1t)$, by injecting the cyber-attack signal (\ref{eq:cyber-attacksignalstatic2}) to the leader's communication channels, it can be seen that the follower vehicles will follow the attacker as demonstrated in Fig. \ref{fig:StatesStaticNonstealthy}. However, as stated in Section \ref{sec:StaticController}, this cyber-attack signal will not be stealthy and will be detected by the follower vehicles in the set $\mathcal{V}_d$ as demonstrated in Fig. \ref{fig:residualinStaticNonstealthy}.


	\section{Conclusion}
	\label{sec:Conclusion}
This paper introduces a comprehensive framework for stealthy attacks in leader-follower vehicle platoons, where adversaries manipulate vehicles to track attacker-desired trajectories while remaining undetected. Unlike existing research focused on destabilizing attacks, our work demonstrates sophisticated manipulation attacks that exploit inter-vehicle communication trust. Our analysis shows attackers can achieve objectives in various scenarios. When all channels are vulnerable, minimal system knowledge suffices but requires leader state access. When leader channels are secure, attackers can still manipulate some subsets of the followers through more sophisticated attacks requiring greater system knowledge.

Future directions include extending analysis to heterogeneous platoons, investigating time-varying topology impacts, developing transient-effective detection mechanisms, and creating integrated privacy-preserving and attack-resilient protocols. This work provides foundation for securing the next-generation connected autonomous vehicle systems and maintaining safety under sophisticated adversarial conditions.

\appendices
\section{Proof of Lemma 1}
\label{Appendix 1}
By denoting $\zeta_i=x_i-x_0+d_i$, constructing $\zeta=\Big[\zeta_{1}^{T},...,\zeta_{N}^{T}\Big]^T$ and $E=diag(e_1,...,e_N)$, we can obtain:
\begin{align}
	\begin{split}
		\dot{\zeta}&=(I_N \otimes A + E\mathcal{L}_1 \otimes BK)\zeta \\&+ (E\otimes B)sgn((\mathcal{L}_1 \otimes K)\zeta)-(1\otimes B)u_0
	\end{split}
\end{align}

Then, consider the following Lyapunov function candidate:
\begin{align}
	V=\zeta^T (\mathcal{L}_1 \otimes P^{-1})\zeta + \sum_{j \in \mathcal{N}_i}\frac{1}{\tau}(e_i-\alpha)^2
\end{align} 
where $\alpha>0$ is a positive constant. Therefore, we obtain:
\begin{align}
	\begin{split}
		\dot{V}&=2\zeta^T(\mathcal{L}_1 \otimes P^{-1}A+\mathcal{L}_1E\mathcal{L}_1 \otimes P^{-1}BK)\zeta \\& + 2\zeta^T (\mathcal{L}_1E \otimes P^{-1}B)sgn((\mathcal{L}_1\otimes K)\zeta)\\&-2\zeta^T(\mathcal{L}_1 1 \otimes P^{-1}B)u_0+\sum_{j \in \mathcal{N}_i}\frac{2}{\tau_i}(e_i-\alpha)\dot{e}_i
	\end{split}
\end{align}
By noting that $K=-B^TP^{-1}$, we have:
\begin{align*}
	&\zeta^T(\mathcal{L}_1E\mathcal{L}_1 \otimes P^{-1}BK)\zeta \\&= -\sum_{j \in \mathcal{N}_i} e_i \Bigg[\sum_{j \in \mathcal{N}_i}(\zeta_i-\zeta_j)\Bigg]^T \\&\times P^{-1}BB^TP^{-1}\Bigg[\sum_{j \in \mathcal{N}_i}\zeta_i-\zeta_j\Bigg]
\end{align*}
Note that we have the following for $\dot{e}_i$:
\begin{align*}
	\dot{e}_i&= \tau_i \Bigg[\sum_{j \in \mathcal{N}_i}\zeta_i-\zeta_j\Bigg]^T \times \Gamma \Bigg[\sum_{j \in \mathcal{N}_i}\zeta_i-\zeta_j\Bigg]\\&+\tau_i||K(\sum_{j \in \mathcal{N}_i}\zeta_i-\zeta_j)||_1
\end{align*}	
Furthermore, we have:
\begin{align*}
	&\zeta^T(\mathcal{L}_1E \otimes P^{-1}BK)sgn((\mathcal{L}_1\otimes K)\zeta)\\&=-\sum_{j \in \mathcal{N}_i}e_i\Bigg[\sum_{j \in \mathcal{N}_i}\zeta_i-\zeta_j\Bigg]^TP^{-1}B \\&\times sgn\Bigg(B^{T}P^{-1}\Big[\sum_{j \in \mathcal{N}_i}\zeta_i-\zeta_j\Big]\Bigg)\\&= -\sum_{j \in \mathcal{N}_i}e_i||B^TP^{-1}\Big[\sum_{j \in \mathcal{N}_i}\zeta_i-\zeta_j\Big]||_1
\end{align*}
Therefore, since $\Gamma=P^{-1}BB^TP^{-1}$ and given the above equations, by considering $\tilde{\zeta}=(I_N\otimes P^{-1})\zeta$, one can obtain:
\begin{align*}
	\dot{V}&=\tilde{\zeta}^T\Big[\mathcal{L}_1 \otimes (AP+PA^T)-2\alpha \mathcal{L}_{1}^{2} \otimes BB^T\Big] \tilde{\zeta}\\&-2\alpha || (\mathcal{L}_1\otimes B^T)\tilde{\zeta} ||_1 - 2\tilde{\zeta}^T(\mathcal{L}\mathbb{1}\otimes B)u_0 \\&
	\le \tilde{\zeta}^T \Big[\mathcal{L}_1 \otimes (AP+PA^T)-2\alpha \mathcal{L}_{1}^{2} \otimes BB^T\Big]\tilde{\zeta} \\&- 2(\alpha-\gamma)||(\mathcal{L}_1 \otimes B^T)\tilde{\zeta}||_1
\end{align*}
Therefore, by taking $\alpha$ to be satisfy $\alpha>\gamma$ and $\alpha \lambda_i>1,\forall i=1,...,N$, we obtain:
\begin{align}
	\label{eq:Theorem4midlYapunov}
	\dot{V}< 2\tilde{\zeta}^T\Big[\mathcal{L}_1 \otimes (AP+PA^T)-2\alpha\mathcal{L}_{1}^{2} \otimes BB^T\Big]\tilde{\zeta}
\end{align}

Then, by letting $U \in \mathbb{R}^{N \times N}$ to be a unitary matrix such that $U^T\mathcal{L}_1 U=\Lambda= diag(\lambda_1,...,\lambda_N)$ and $\zeta = (U^T \otimes I_n)\zeta$. Then, we have:
\begin{align}
	\begin{split}
		&\mathcal{L}_1 \otimes (AP+PA^T)-2\alpha\mathcal{L}_{1}^{2} \otimes BB^T \\&=
		\sum_{i=1}^{N}\lambda_i (AP+PA^T - 2\alpha \lambda_iBB^T)
	\end{split}
\end{align}
Therefore, given that $\alpha \lambda_i >1$ and the inequality (\ref{eq:Ricatti}), we get $\dot{V}< 0$. Therefore, $\zeta$ will be asymptotically stable, indicating that the follower vehicles will track the leader's state while maintaining the required distance with the leader. This completes the proof of the Lemma. \hfill $\blacksquare$
\section{Proof of Theorem 1}
\label{appendix:2}
By denoting $\delta_i=x_i-x_a+d_i$, we can then construct $\delta=\Big[\delta_{1}^{T},...,\delta_{N}^{T}\Big]^T$ and $E=diag(e_1,...,e_N)$. Then, we can obtain:
\begin{align}
	\begin{split}
		\dot{\delta}&=(I_N \otimes A + E\mathcal{L}_1 \otimes BK)\delta \\&+ (E\otimes B)sgn((\mathcal{L}_1 \otimes K)\delta)-(1\otimes B)u_a
	\end{split}
\end{align}

Then, by considering the following Lyapunov function candidate:
\begin{align}
	V=\delta^T (\mathcal{L}_1 \otimes P^{-1})\delta + \sum_{j \in \mathcal{N}_i}\frac{1}{\tau}(e_i-\alpha)^2
\end{align} 
where $\alpha>0$ is a positive constant, we can follow the steps similar to Lemma 1 to show that $\delta$ will be asymptotically stable, indicating that the vehicles will track the attacker's state while satisfying the required spacing.

We now turn our focus to the stealthiness of the attack signal (\ref{eq:CyberAttackDynamic1}). We can obtain the following:
\begin{align*}
	&||x_i - x_{0}^{c}+d_i||=||x_i-x_0-x_a+x_0+d_i||=\\&||x_i-x_a+d_i||=||\delta_i||=0
\end{align*}
Hence, the attack (\ref{eq:CyberAttackDynamic1}) will be stealthy. This completes the proof of the Theorem. \hfill $\blacksquare$

\section{Proof of Theorem 2}
\label{appendix:3}
By considering the attacker as a virtual node, acting as the root node for the vehicles in the set $\mathcal{V}_{nd}$, one can obtain the following:
\begin{align*}
	\dot{\delta}_j&=A\delta_j + Be_jK\sum_{k \in \mathcal{N}_{j}^{\mathcal{V}_{nd}}}(\delta_j-\delta_k)\\&+e_jsgn\Big(K\sum_{k \in \mathcal{N}_{\mathcal{V}_{nd}}}\delta_j-\delta_k\Big)-Bu_a
\end{align*}
where $j \in {\mathcal{V}_{nd}}$. One can then follow the steps similar to proof of Theorem 1 to show that the vehicles in this set will track the attacker while satisfying the spacing requirements, i.e., $x_j=x_a-d_{j}^{c}, \forall j \in {\mathcal{V}_{nd}}$. Then, the information that the vehicles in the set ${\mathcal{V}_{d}}$ receive from the vehicles in the set ${\mathcal{V}_{nd}}$ is as follows:
\begin{align*}
	x_{j}^{c}=x_j+a_{ji}=x_j+x_0-d_j-x_a+d_{j}^{c}=x_0-d_j
\end{align*}
Therefore, since due to Lemma 1, the agents in the set $\mathcal{V}_d$ are tracking the leader with the spacing requirements, i.e., $x_i=x_0-d_i$, to show stealthiness of the attack, we have:
\begin{align*}
	x_{i}-x_{j}^{c}+d_{ij}&=x_i-x_0+d_j+d_{ij}\\&=x_i-x_0+d_i\\&=\zeta_i=0
\end{align*}
Therefore, as $t \to \infty$, $r_{x}^{i}=0$. Furthermore, the residual for the agents in the set $\mathcal{V}_{nd}$ will be:
\begin{align*}
	x_{j}-x_{i}^{c}+d_{ji}&=x_j-x_{i}-x_{a}+x_0-d_i+d_{ji}\\&=
	x_j-x_a+d_{ji}=0
\end{align*}
where we have utilized the fact that $d_{j}^{c}=d_{ji}$, i.e., the spacing between the attacker and vehicle $j$ will be equal to the spacing between vehicle $j$ and $i$ in attack free scenarios. Therefore, $r_{x}^{j}=0$ as $t \to \infty$.
This completes the proof of the Theorem. \hfill $\blacksquare$

\section{Proof of Theorem 3}
\label{Appendix:4}
The steps of the proof is similar to Theorem 2. Therefore, similar to Theorem 2, the vehicles in the set $\mathcal{V}_{nd}^{a}$ will track the attacker while satisfying the spacing requirements.

In order to show stealthiness, first note that for the vehicles in the set $\mathcal{V}_{nd}^{c}$, the proof of stealthiness is similar to Theorem 2. For the vehicles in the set $\mathcal{V}_{nd}^{dc}$, since these vehicles will not track the leader due to a lack of connectivity with the leader, one can construct the following in the under-attacked vehicles to demonstrate stealthiness:
\begin{align*}
	x_{j}-x_{i}^{c}+d_{ji}&=x_{j}-x_i-a_{ij}+d_{ji}\\&
	=x_{j}-x_i-x_a+x_i+d_{j}^{c}+d_{ij}+d_{ji}\\&=
	x_a-d_{j}^{c}-x_i-x_a+x_i+d_{j}^{c}+d_{ij}+d_{ji}\\&=0
\end{align*}
where $i \in \mathcal{V}_{nd}^{dc}$ and $j \in {\mathcal{V}_{nd}^{a}}$. Furthermore, for the vehicles in $\mathcal{V}_{nd}^{dc}$, we have:
\begin{align*}
	x_i-x_{j}^{c}+d_{ij}&=x_i-x_{j}-a_{ji}+d_{ij}\\&
	=x_i-x_j-x_i+x_a+d_{ji}+d_{ij}-d_{j}^{c}\\&
	=0
\end{align*}
where $i \in \mathcal{V}_{nd}^{dc}$ and $j \in {\mathcal{V}_{nd}^{a}}$ due to the fact that $x_j=x_a-d_{j}^{c}$ and $d_{ij}=-d_{ji}$.
This complete the proof of the Theorem. \hfill $\blacksquare$

\section{Proof of Theorem 4}
\label{Appendix:5}
 We first demonstrate that the agents track the attacker and then, we show that the attack (\ref{eq:AttackDesignStatic}) is stealthy based on Definition 2. Let us construct the tracking error $\delta_i=x_i-x_a+d_i$, we have:
\begin{align}
	\begin{split}
		\dot{\delta}_i&=A\delta_i + c_1 BK \sum_{j \in \mathcal{N}_i}(\delta_i - \delta_j) \\&+ c_2Bsgn(K\sum_{j \in \mathcal{N}_i}(\delta_i - \delta_j)) - Br_a
	\end{split}
\end{align}
Therefore, by letting $\delta=[\delta_{1}^{T},...,\delta_{N}^{T} ]^T$, we obtain:
\begin{align}
	\label{eq:Theorem1mid1}
	\begin{split}
		\dot{\delta}&=(I_N \otimes A + c_1 \mathcal{L}_1 \otimes BK)\delta \\&+ c_2 (I_N \otimes B)sgn((\mathcal{L}_1 \otimes K )\delta)-(\mathbf{1} \otimes B)u_a
	\end{split}
\end{align}
Therefore, take the following Lyapunov function candidate:
\begin{align}
	V=\zeta^T (\mathcal{L}_1 \otimes P^{-1}) \zeta
\end{align}
Then, the proof that the follower vehicles follow the attacker can be obtained by following the steps provided in the proof of Theorem 1 in \cite{li2012distributed}.

Furthermore, proof of stealthiness is also similar to Theorem 1 and is omitted for the sake of brevity. This completes the proof of this Theorem. \hfill $\blacksquare$

\section{Proof of Theorem 5}
\label{Appendix:6}
One can write the tracking error $\delta_i=x_i-x_a-d_i$ for all the follower vehicles as follows:
\begin{align}
	\begin{split}
		\dot{\delta}_i&=A\delta_i + c_1 BK \sum_{j \in \mathcal{N}_i}(\delta_i - \delta_j) \\&+ c_3Bsgn(K\sum_{j \in \mathcal{N}_i}(\delta_i - \delta_j)) - Br_a
	\end{split}
\end{align}
Then, with an analysis similar to Theorem 1, one can obtain:
\begin{align}
	\label{eq:Lyapunovmid3}
	\begin{split}
		\dot{V} &\le \delta^{T} (\mathcal{L}_1 \otimes (AP+PA^T)-2c_1 \mathcal{L}_{1}^{2} \otimes BB^T)\delta \\&- 2(c_3 - \gamma_2)|| (\mathcal{L}_1 \otimes B^T)\delta||_1
	\end{split}
\end{align}
The rest of the proof will be similar to Theorem 1 and Theorem 1 of \cite{li2012distributed}. Note that different from Theorem 1, we have $\gamma_2 \ge \gamma$, which demonstrate that the attackers are not bounded by the upper bound of the leader's input signal $u_0$. 

In order to show that the attack signal will not be stealthy, One can observe that  $||x_i - x_{0}^{c}+d_{ij}||=||x_i-x_0-x_a+x_0+\frac{c_3-c_2}{c_1 ||K||}sgn(K\sum_{j\in \mathcal{N}_i }(x_i-x_j))+d_{ij}||=||x_i-x_a+d_{ij}+\frac{c_3-c_2}{c_1 ||K||}sgn(K\sum_{j\in \mathcal{N}_i }(x_i-x_j+d_{ij}))||=||\frac{c_3-c_2}{c_1 ||K||}sgn(K\sum_{j\in \mathcal{N}_i }(x_i-x_j+d_{ij}))||$. Therefore, since $r_{x}^{i} > 0$, this attack will be detected. This completes the proof of the Theorem. \hfill $\blacksquare$

\bibliographystyle{IEEEtran}
\bibliography{sample}  

\end{document}